# Nonlinear Dynamics of Capacitive Charging and Desalination by Porous Electrodes

P.M. Biesheuvel[1,2] and M.Z. Bazant[3]

[1]*Department of Environmental Technology, Wageningen University, Bomenweg 2, 6703 HD Wageningen, The Netherlands.* [2]*Wetsus, centre of excellence for sustainable water technology, Agora 1, 8900 CC Leeuwarden, The Netherlands.* [3]*Departments of Chemical Engineering and Mathematics, Massachusetts Institute of Technology, Cambridge, MA 02139, USA.*

**Abstract**

The rapid and efficient exchange of ions between porous electrodes and aqueous solutions is important in many applications, such as electrical energy storage by super-capacitors, water desalination and purification by capacitive deionization (or desalination), and capacitive extraction of renewable energy from a salinity difference. Here, we present a unified mean-field theory for capacitive charging and desalination by ideally polarizable porous electrodes (without Faradaic reactions or specific adsorption of ions) in the limit of thin double layers (compared to typical pore dimensions). We illustrate the theory in the case of a dilute, symmetric, binary electrolyte using the Gouy-Chapman-Stern (GCS) model of the double layer, for which simple formulae are available for salt adsorption and capacitive charging of the diffuse part of the double layer. We solve the full GCS mean-field theory numerically for realistic parameters in capacitive deionization, and we derive reduced models for two limiting regimes with different time scales: (i) In the "super-capacitor regime" of small voltages and/or early times where the porous electrode acts like a transmission line, governed by a linear diffusion equation for the electrostatic potential, scaled to the RC time of a single pore. (ii) In the "desalination regime" of large voltages and long times, the porous electrode slowly adsorbs neutral salt, governed by coupled, nonlinear diffusion equations for the pore-averaged potential and salt concentration.

## I. Introduction

Porous electrodes in contact with aqueous solutions[1-4] are found in many technological applications, such as the storage of electrical energy in (electrostatic double-layer) super-capacitors,[5-9] capacitive deionization (CDI) for water desalination,[2,10-17] and the reverse process of CDI, namely the capacitive extraction of energy from the salinity difference between different aqueous streams, for instance river and sea water.[18] In all of these cases, the porous electrodes can be assumed to be ideally polarizable, i.e. blocking to Faradaic electron-transfer reactions (in contrast to porous electrodes used in batteries and fuel cells). Over the past five years, the nonlinear dynamics of capacitive charging for locally flat, ideally polarizable electrodes and metallic particles have been extensively studied,[19-24] but the different geometry of porous electrodes has not been yet analyzed in comparable detail. In this paper, we develop a unified mean-field theory for the nonlinear charging dynamics of ideally polarizable, porous electrodes and apply it to the model problem sketched in Fig. 1.

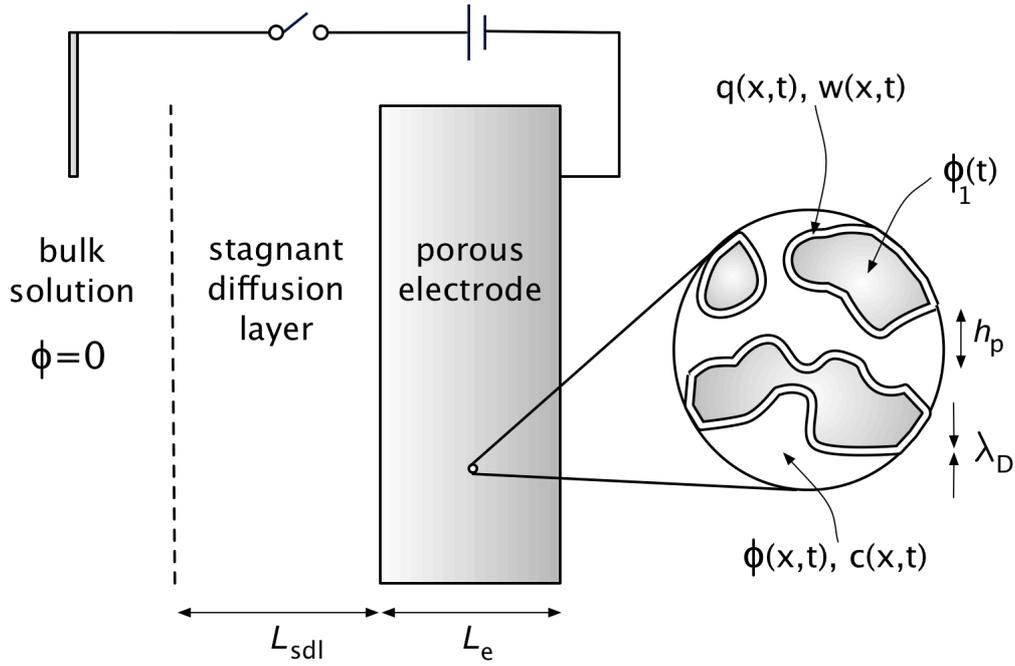

Figure 1. Sketch of the model problem. At $t=0$, a voltage $\phi_1$ is applied to an ideally polarizable, porous electrode of thickness $L_e$, relative to a bulk electrolytic solution separated by a stagnant diffusion layer of thickness $L_{sdl}$. The characteristic pore thickness $h_p$ (defined as the ratio of the pore volume to the pore area) is much larger than the Debye screening length, $\lambda_D$, so the pore space is mostly filled with quasi-neutral electrolyte, exchanging ions with a charged, thin double-layer "skin" on the electrode matrix. The volume-averaged potential $\phi(x,t)$ and neutral salt concentration $c(x,t)$ vary in space and time as counterions enter, and co-ions leave, the thin double layers, thus changing the mean surface charge density $q(x,t)$ and total excess salt density $w(x,t)$.

Super-capacitors store electrical energy by the physical adsorption of counter-ions in high-surface-area electrochemical double layers within a porous electrode. Assuming long, straight electrode pores with thin double layers, ion transport can be approximated by a linear "RC" transmission line,[7,25] where the neutral solution in the pore acts like a core resistance and the double layer like a coaxial capacitance, enclosed in a coaxial electron-conducting sheath, see Fig. 2. This equivalent-circuit model is still widely used to describe the linear response of super-capacitor electrodes, also with more complicated internal geometries.[26,27] The possibility of nonlinear response due to large applied voltages and/or narrow pores (thick double layers), leading to local depletion of ions, has received much less attention.

CDI is a desalination (ion-removal) process in which an aqueous solution flows through the space in between two porous electrodes. Upon applying a voltage difference between the two electrodes, cations are transported to the electrode of negative bias, and anions to the other. These ions are stored as counter-ions within the structure of the porous electrodes. Simultaneous with counter-ion adsorption, co-ions are expelled from the electrodes, but co-ion outflow is always less than counter-ion adsorption. As a consequence of the resulting net ion removal, the product solution becomes partially depleted in ions and in this way for instance potable water (approx. <8 mM ionic strength) can be produced from brackish water (>15 mM). When the equilibrium ion adsorption capacity of the electrodes has been reached, the voltage difference can be reduced, ions are released, and a flow concentrated in salt is produced, after which the deionization cycle can be repeated. Transmission line models have also been applied to CDI at low voltages,[2] but for practical desalination systems, it is crucial to predict the nonlinear response of porous electrodes, significantly altering the bulk ionic strength.

To describe ion transport in an electrode by a mean-field theory we must consider the two interpenetrating phases, first the electron-conducting electrode matrix, and second the aqueous phase which fills the pores in between the electrode matrix. At the interface of these two phases, the electrode internal matrix-solution surface, the electron charge is locally compensated by an excess ion charge, i.e., a double layer is formed. The difference in voltage between the conducting matrix, $\phi_1$ ($\Phi_1$ in Newman's[1-4] terminology), and that of the aqueous solution in the pores, $\phi$ ($\Phi_2$), is given by the electrostatic potential difference across this interfacial double layer.

To keep the mathematical model tractable we will consider dilute solution theory, without applying corrections for ion crowding and dielectric saturation.[28,29] In particular, for a dilute, binary electrolyte, we can then use the Gouy-Chapman-Stern (GCS) model in which the double layer is described by a combination of a charge-free Stern layer of constant capacity (other names for the Stern layer are: inner, compact, or Helmholtz layer), and a diffuse layer. The potential difference $\phi_1$-$\phi$ equals the potential difference across the Stern layer, $\Delta\phi_S$, plus the potential difference across the diffuse part of the double layer, $\Delta\phi_D$. The electrical resistance in the conducting matrix is typically much smaller than the ionic resistance in the pores, and therefore we can assume a constant voltage within the conducting matrix. After applying an electrode potential $\phi_1$ (e.g., relative to another porous electrode, such as is the case in the aforementioned processes), the potential within the aqueous solution phase within the pores of the electrode, $\phi$, initially remains equal to that in the electrode matrix, $\phi_1$, because double layers have not yet formed, while the applied potential difference between the two electrodes decays across the aqueous solution in between the two electrodes. The field strength outside the electrode induces a differential flow of ions into and out of the electrodes and consequently leads to charge formation at the internal matrix/solution-interface within the electrode. After sufficient time, and for a purely capacitive process without electrochemical charge transfer, the applied voltage difference will be fully compensated in the double layers at the matrix/solution-interfaces, and the potential in the pores $\phi$ becomes constant across the electrode and equal to the potential within the solution phase outside the electrodes, i.e., equilibrium will be reached.

In the present work, a unified mean-field theory is presented to describe ion transport and storage in a porous electrode. Only one electrode is considered, and calculation results are presented for a prescribed voltage difference between electrode and bulk solution. To simplify the calculations, a constant ion concentration is assumed at a point in the bulk solution, far from the electrode. Of course, in some of the applications, such as in CDI, the goal is to strongly modify the bulk salt concentration, but here we focus on modeling the dynamics of ions within the porous electrode, which in a later stage can be combined with more complicated models of the bulk solution, e.g. allowing for fluid flow and other geometries. As noted above, we also neglect Faradaic charge transfer reactions and non-electrostatic (specific) ion adsorption. Instead, only physical adsorption in the diffuse part of the double layer is considered to compensate the electron charge. Because Faradaic reactions are not included, the present calculation describes a purely capacitive process, where the steady state is characterized by zero current and vanishing ionic fluxes.

An important point to realize is that to describe the charge separation at the matrix/solution-interface (according to the diffuse double layer model) it is not sufficient to only consider the (differential) capacity, i.e., the (differential) relation between the difference $\phi_1$-$\phi$ and the local charge storage $q$. Instead at the same time we must describe how many salt ions are adsorbed in the double layer. Though in a hypothetical calculation in which only counter-ions are considered to adsorb (as in the Helmholtz model for the double layer), charge density and salt adsorption are equal (because the electron density is exactly compensated by the density of counter-ion charge), this is not the case when ions are assumed to be stored diffusively in the Gouy-Chapman part of the double layer. In this case, the electron charge -$q$ is compensated both by the accumulation of counter-ions, as well as by the expulsion from the diffuse part of the double layer of co-ions. For low potentials across the diffuse layer (the Debye-Hückel limit) both contributions are of equal importance, i.e., for each electron stored, there is half a cation adsorbed and half an anion desorbed, while only for very high potentials the limit is slowly approached that counterion adsorption fully compensates the electron charge. This effect can be quantified by making use of the charge efficiency, $\Lambda$, which is the ratio of net salt adsorption $w$ over charge, $q$, which in the GCS-model changes from zero at low voltage to unity at high voltage. Recently, it was shown that data for charge efficiency, based on equilibrium data for salt adsorption and charge of porous electrodes as function of voltage and salt concentration, could be very well explained by the GCS-model, providing support for the validity of this model to describe the structure of the double layer in porous electrodes.[30]

The transport equations we present are similar to those developed several decades ago by J. Newman and collaborators[1-4] to describe ion transport in porous electrodes. Since then, however, it seems that the problem of ion transport within the pores of an electrode, coupled to transient charging of the double layers at the internal electrode surface, has not yet been satisfactorily solved, at least not in the context of mean-field porous electrode theory. The analysis of Newman and Tiedemann[3] requires solution of their expression for the differential of surface concentration. However, as they write, "This formidable equation indicates all the differential coefficients which must be known for an exact treatment of transient processes involving double layer charging. … The general case involving double-layer charging and all the differential coefficients becomes too complicated to pursue profitably at this point." To the best of our knowledge, since that time, a complete mathematical theory of capacitive charging and desalination by porous electrodes has not yet emerged.

The time is ripe to tackle this outstanding theoretical challenge, since much progress has been recently made on analogous problems of nonlinear electrochemical relaxation around smooth (locally flat) polarizable surfaces. Bazant, Thornton, and Ajdari[19] (BTA) first analyzed the transient nonlinear response to a suddenly applied voltage step across parallel-plate electrodes (using matched asymptotic expansions in the limit of thin double layers) and identified two distinct dynamical regimes: (i) The "weakly nonlinear" regime of small applied voltages or early times, when the system behaves like an RC circuit, with the quasi-neutral bulk resistance in series with the double layer capacitors, and

(ii) the "strongly nonlinear" regime of large applied voltages and long times, when ion adsorption by the double layers depletes the local salt concentration and slaves the charging process to the slow arrival of additional ions by bulk diffusion. A central aspect of this work[19] was to quantify, apparently for the first time, the coupled, nonlinear effects of capacitive charging and neutral salt adsorption by thin double layers (Eqs. 137-139, respectively, corresponding to the GCS model, derived by asymptotic flux matching). The BTA analysis has since been extended by many authors, e.g to polarizable particles in step electric fields[20] (also taking into account tangential ion transport through the double layers[31]), ionic liquids in large step voltages[21] (which exhibit additional dynamical regimes), electrolytes in large AC voltages[23,24] (accounting for the imposed time scale and the transient formation of extended space charge[32]), and AC electro-osmotic flows[28,33] (which couple ionic relaxation to fluid motion[34]). In all of these situations, the same two dynamical regimes can be identified, separated by a transition voltage (around 10 $kT/e$) where neutral salt adsorption by the double layers begins to dominate capacitive charging.

In this paper, we extend this work to the fundamentally different geometry of porous electrodes, motivated by applications in energy storage and desalination, which were not even mentioned by BTA or subsequent authors (whose focus was on micro-electrodes and colloids). We identify two analogous dynamical regimes, in which a porous electrode behaves either as a super-capacitor (weakly nonlinear response) or as a desalination system (strongly nonlinear response). A key quantity controlling this dynamical transition is the charge efficiency ($w/q$ in the notation of BTA), as co-ion expulsion at low voltages is replaced by additional counter-ion adsorption at high voltages. We illustrate these basic principles by numerical simulations of CDI using the full, nonlinear mean-field theory and by analyzing reduced model equations for the limiting regimes.

**II. Theory**

We begin by presenting our simple mathematical framework, which describes the nonlinear dynamics of ion transport within a porous electrode and the adjacent solution phase, without any *ad hoc* assumptions of local steady state in either region. The key assumption is that of thin double layers compared to the typical pore size in the electrode, so that there is a clear distinction between the quasi-neutral solution in the center of a pore and its surrounding thin double-layer "skin", containing diffuse ionic charge, screening the surface charge[3,19,20,35-39]. This classical approximation can be justified by matched asymptotic expansions, even in time-dependent problems with curved surfaces[19,31]. We also neglect tangential surface transport through the double layers compared to quasi-neutral bulk transport within a pore, which is valid for thin double layers, as long as the surface charge is sufficiently small (Dukhin/Bikerman number<<1).[20,31] This ubiquitous assumption becomes violated in the "strongly nonlinear" regime,[19] where double-layer salt adsorption significantly depletes the salt concentration at the pore center, and more general, nonlinear "surface conservation laws" should be used as effective boundary conditions on the quasi-neutral solution within a pore.[20,31] As a first approximation for nonlinear dynamics, however, we follow all prior work on porous electrodes and neglect tangential surface conduction within the pores. Compared to smooth electrodes and particles, effects of surface conduction may be reduced in typical porous microstructures, due to random surface orientations, and it may be possible to account for such effects heuristically in the effective transport properties of the porous electrode.

Our approach is based on J. Newman's macroscopic porous electrode theory,[1-4] and we make the same classical and widely used assumptions. (See *Appendix A* for a formal derivation from microscopic transport equations within the pores.) The local salt concentration and electrostatic potential of the quasi-neutral solution within the pores are assumed to vary slowly enough to permit volume averaging to yield smooth macroscopic variables. The exchange of ions with the double layers on the pore surfaces is modeled as a slowly varying volumetric source/sink term in the macroscopic, volume-averaged transport equations. The porous electrode is thus treated as a homogeneous mixture of charged double layers and quasi-neutral solution, regardless of the geometrical structure of the pores.

When ions are transported from solution into a porous electrode under the influence of an applied electrical potential difference, ion concentration and potential gradients will also develop in the bulk solution outside the electrode. To approximate and simplify the description of ion transport in solution, we make the standard engineering approximation of a mass transfer film adjacent to the electrode. This layer goes under many names such as the Nernst layer, concentration-polarization layer, advection-diffusion layer, or just diffusion layer, and is a well-known concept in electrochemical process modeling and in the field of charged (electro-dialysis) membranes. Here we will denote it as the "stagnant diffusion layer" (SDL) and consider it to have a constant thickness, which is applicable to a process in which there is a certain extent of convective mixing, which limits the diffusive spreading of the diffusion layer into the bulk solution.

The assumptions above suffice to develop a complete model for any electrolytic solution. Here, we illustrate the basic principles for the canonical case of a symmetric binary $z$:$z$ electrolyte with equal free solution diffusivity for cations and anions, $D$. Within the pores of the electrodes the diffusivity, $D_e$, will typically be lower than in solution, but again we take the same value for an- and cation. This is an effective axial diffusivity which includes effects of pore tortuosity and pore wall friction. With only two ions present, quasi-neutrality implies equal concentrations $c_+=c_-=c$, where $c(x,t)$ is the neutral salt concentration. Neglecting various non-ideal concentrated-solution effects,[4,28,29] the Nernst-Planck (NP) equation for a dilute solution expresses the flux of ionic species i as a sum of contributions from diffusion and electromigration. In dimensionless form, the NP equation for the pore phase within the electrode can be written as

$$j_i = -\tfrac{1}{2}\left(\nabla c_i + z_i c_i \nabla \phi\right) \tag{1}$$

where $c_i$ is a dimensionless ion concentration, $c_i=C_i/C_\infty$, where $C_\infty$ is a reference ionic strength, e.g., that of the bulk solution on the outside of the SDL, $x$ is the dimensionless position, $x=X/L_e$ with $L_e$ the thickness of the electrode, and $\phi$ is the dimensionless electrostatic potential in the pores, which becomes dimensional after multiplication with the thermal voltage, $V_T=k_B T/ze$. In Eq. (1) the dimensionless ion flux $j_i=J_i/J_{lim}$ is scaled to the diffusion-limited current $J_{lim}=2\cdot D_e\cdot C_\infty/L_e$. Note that we define concentrations, currents, etc., within the electrode based on the open area, i.e., based on the area accessible to the ions. Eq. (1) is the basis for the further theory, with $j_++j_-$ the net salt flow into/out of the electrode, and with $j_+-j_-$ equal to the axial current, $i_e$.

The salt balance in the electrode pores is equivalent to Fick's second law extended to include the local rate of salt adsorption, $j_{salt}$ (=$J_{salt}/J_{lim}$), from the pore solution into the double layer at the matrix/solution-interface,

$$\frac{\partial c}{\partial t} = \nabla^2 c - \alpha\, j_{salt} \qquad (2)$$

where $c=c^+=c^-$ is the dimensionless salt concentration, $t$ is a dimensionless time, obtained by rescaling the dimensional time $\tau$ with $\tau_d = L_e^2/D_e$. The parameter $\alpha$ is a dimensionless specific surface area density of the solution/matrix-interface, given by $\alpha = a \cdot L_e/p$, where $a$ is a dimensional surface density in area per total electrode volume, which has units of inverse length, and $p$ is the porosity of the electrode. It can also be expressed as the ratio, $\alpha = L_e/h_p$, of the electrode length to a characteristic pore thickness, $h_p = p/a$, defined as the ratio of the pore volume to the pore surface area. (For a cylindrical pore, $h_p$ is half of the pore radius.) Equation (2) describes the variation of concentration $c$ with depth $x$ in the electrode (axial direction) by transient diffusion and volume-averaged salt removal $j_{salt}$ at each position from the aqueous phase and to the diffuse double layers on the electrode surface.

Coupled to Eq. (2), we must solve in the pore phase of the electrode a differential Ohm's law for the current carried by the ions, $i_e$, namely

$$i_e = -c\,\nabla\phi. \qquad (3)$$

Within the electrode the current carried by the ions, $i_e$, is not constant but will decrease with depth within the electrode (to reach zero at the backside of the electrode), at each point compensated by the electron current in the matrix (which increases from zero at the SDL/electrode-interface to a maximum at the backside of the electrode), such that the total current remains the same at each position. Furthermore a local charge balance describes how the ion current $i$ in the electrode decreases with depth because charge is transferred to the double layers at the solution/matrix-interface,

$$\nabla \cdot i_e = -\alpha\, j_{charge}. \qquad (4)$$

This finalizes the macroscopic description of transport within the aqueous phase of the electrodes.

Next we describe the double layer model which is solved at each position $x$ within the electrode and which locally describes the potential difference between matrix phase and solution, $\phi_1-\phi$, which is equal to the sum of potentials across the Stern and diffuse parts of the double layer model, thus $\phi_1-\phi=\Delta\phi_S+\Delta\phi_D$. We base the description of charge formation at the matrix/solution-interface on the Gouy-Chapman-Stern model, which combines a Stern layer of constant capacity, with a diffuse, Gouy-Chapman, layer. The GCS-model is based on the assumption of ions as ideal point-charges in local thermodynamic equilibrium within a planar, isolated interface. Two equations suffice for the GCS-model. The first is

$$q = -2\beta\sqrt{c}\,\sinh\frac{\Delta\phi_D}{2} \qquad (5)$$

where $q$ is a dimensionless surface charge density (of the excess charge in the diffuse layer; multiply $q$ by $\lambda_B$, $C_\infty$, $N_{av}$ and the electron charge, $e$, to obtain a dimensional surface charge density), $\beta = 2\lambda_D^0/\lambda_B\,c$, $c$ is the dimensionless ion concentration in the pore at position $x$, and $\Delta\phi_D$ is the potential difference over the diffuse layer. At room temperature, the Bjerrum length $\lambda_B$ is ~0.72 nm. The Debye length $\lambda_D^0 = \kappa^{-1}$ relates to $C_\infty$ according to $\kappa^2 = 8\pi \cdot \lambda_B \cdot N_{av} \cdot C_\infty$. The second GCS equation relates the voltage difference across the Stern layer $\Delta\phi_S$ to $q$ as

$$q = -\frac{\beta \Delta\phi_S}{\delta} \tag{6}$$

where $\delta = \lambda_S / \lambda_D^0$ and $\lambda_S$ is an effective thickness of the Stern layer.[19,36,38,39] There is also a simple formula for the salt adsorption[16,19]

$$w = 4\beta\sqrt{c}\, \sinh^2 \frac{\Delta\phi_D}{4} \tag{7}$$

similar to Gouy's formula for the surface charge density, Eq. (5).

To close the porous electrode model, we invoke mass conservation to relate the volume-averaged rate of charge removal from the electrolyte phase, $j_{charge}$, to the charge density, $-q$, according to

$$\frac{\partial q}{\partial t} = \frac{2 L_e}{\lambda_B} j_{charge} \tag{8}$$

and $j_{salt}$ to the adsorbed salt density, $w$, as

$$\frac{\partial w}{\partial t} = \frac{2 L_e}{\lambda_B} j_{salt}. \tag{9}$$

These are essentially volume-averaged forms of the BTA surface conservation laws, as shown in *Appendix A*.

We have expressed the general theory in terms of multi-dimensional gradient operators applicable to any macroscopic geometry. It is straightforward to integrate the equations over the electrode volume to obtain global conservation laws for ions. In the case of our one-dimensional calculations below, we can check the accuracy of our numerical methods by verifying global charge conservation,

$$\int_{t_1}^{t_2} i_e \big|_{x=0}\, dt = \frac{\lambda_B}{2h_p} \left[ \int_{x=0}^{x=1} q\, dx \right]_{t_1}^{t_2} \tag{10}$$

obtained by integrating Eqs. (4) and (8) in space and time, which equates the time-integrated ion current $i_e$ crossing the SDL/electrode-interface with the change in the total accumulated charge in the double layers integrated over the electrode. Similarly, by integrating Eqs. (2) and (9) in space and time, we obtain the global salt balance,

$$-\int_{t_1}^{t_2} \nabla c \big|_{x=0}\, dt = \left[ \int_{x=0}^{x=1} \left( \frac{\lambda_B}{2h_p} w + c \right) dx \right]_{t_1}^{t_2} \tag{11}$$

which equates the time-integrated diffusive flux crossing the SDL/electrode-interface to the total salt adsorption by the electrode. This concludes the porous-electrode model for purely capacitive response in the absence of any electrochemical processes, such as Faradaic reactions, specific adsorption of ions, or solid intercalation of ions.

In the SDL the salt concentration $c(x,t)$ is described by

$$\frac{\partial c}{\partial t} = d_{sdl} \nabla^2 c \tag{12}$$

where $d_{sdl}=D/D_e$, the ratio in effective diffusivities between solution (sdl) and electrode. The ion current, $i$, is constant across the SDL and based on $i=j_+-j_-$ and $c=c^+=c^-$ follows directly from Eq. (1) as

$$i_{sdl} = -d_{sdl}\, c\, \nabla\phi.  \tag{13}$$

This completes the description of the SDL. Boundary conditions at the SDL/electrode-interface are as follows. At the interface we have continuity in concentration $c_{SDL}=c_e$ and potential, $\phi_{sdl}=\phi_e$. We will include the fact that the electrode is not fully accessible to the aqueous solution and the ions, i.e., the porosity, $p$, is lower than unity. Then, the ion current $i$ on either side is the same but for the porosity correction, thus $i_{sdl}=p\cdot i_e$. Similarly we have continuity in salt flux, which we implement as $d_{sdl}\nabla c\big|_{SDL} = p\,\nabla c\big|_e$. These are the four boundary conditions that apply at the SDL/electrode-interface. At the 'backside' of the electrode, which is blocking for ions, we have $\nabla c = 0$ and $i_e$=0.

### III. Analysis

**III.A. Nonlinear PDEs for the Concentration and Potential**

The system of equation (1)-(9) is highly nonlinear and may appear quite daunting at first, but many variables can be eliminated to obtain a simple pair of partial differential equations (PDEs), e.g. for the volume-averaged salt concentration and electrostatic potential in the electrode. If we neglect the Stern layer capacitance, then the two PDEs can be expressed in an elegant dimensionless form:

$$2\epsilon \frac{\partial}{\partial t}\left[\sqrt{c}\,\sinh\left(\frac{\phi-\phi_1}{2}\right)\right] = \nabla\cdot(c\nabla\phi) \tag{14a}$$

$$4\epsilon \frac{\partial}{\partial t}\left[\sqrt{c}\,\sinh^2\left(\frac{\phi-\phi_1}{4}\right)\right] = \nabla^2 c - \frac{\partial c}{\partial t} \tag{14b}$$

in terms of only one dimensionless parameter,

$$\epsilon = \frac{\lambda_D^0}{h_p} = \frac{\lambda_D^0\, a}{p}. \tag{15}$$

This familiar ratio measures the thinness of the double layers[19,38] by comparing the initial Debye screening length, $\lambda_D^0 = \lambda_D(c=1)$, to the characteristic pore thickness, $h_p = p/a$. The two nonlinear PDEs (14) can also be cast in terms of different pairs of volume-averaged variables, such as the concentrations of cations and anions, the double-layer charge density and neutral salt concentration, etc.

Equation (14)a relates the volume-averaged Maxwell displacement current density (charging the double layers) to the divergence of the ionic current. Equation (14)b relates the volume averaged salt adsorption by the double layers to a sink of diffusing salt in the quasi-neutral solution. As noted above, the theory is based on the assumption of thin double layers in the pores, $\epsilon \ll 1$, and this leads to widely separated time scales, since the small parameter $\epsilon$ multiples one time derivative, and not the other. Mathematically, it is clear from Eq. (14) that there are two different dynamical regimes, with time scales $t = O(\epsilon)$ and $t = O(1)$, determined by the possible dominant balances. Physically, these regimes correspond to different modes of operation of the porous electrode, for either *charge storage* (as a super-capacitor) at early times or *salt removal* (capacitive deionization) at late times, as we now explain. These regimes are completely analogous to those identified and analyzed by BTA[19] for flat electrodes, based on asymptotic analysis involving the same small parameter, $\epsilon \ll 1$, although the geometry of a porous electrode leads to different dynamical equations and scalings.

**III.B. Supercapacitor Regime**

For early times, $t = O(\epsilon)$, there is a dominant balance in Eq. (14)a, which corresponds to capacitive charging of the double layers. Since we start from an initial condition of uniform concentration, $c(x,0)=1$, Equation (14)b representing salt diffusion can be neglected at first, so the quasi-neutral solution phase has an initially uniform conductivity. Concentration variations then appear as small corrections, which can be calculated perturbatively. (The range of validity of this approximation is discussed in the next section.) During this regime, the double layer voltage drop is initially small, $|\Delta\phi| = |\phi - \phi_1| \ll 1$ (or $\ll kT/ze$ with dimensions), since we assume initially uncharged double layers in order to focus only on the charge induced by the applied voltage.

To analyze the earliest times of our model problem, we expand the PDEs for small voltages to obtain a linear initial-boundary-value problem for 0<x<1 (inside the electrode) based on a diffusion equation for the potential,

$$\frac{\partial \phi}{\partial \tilde{t}} = \frac{\partial^2 \phi}{\partial x^2}, \quad \phi(x,0) = \phi_1, \quad \frac{\partial \phi}{\partial x}(1,\tilde{t}) = 0, \quad \frac{\partial \phi}{\partial x}(0,\tilde{t}) = \text{Bi}_e \cdot \phi(0,\tilde{t}) \tag{16}$$

with a rescaled time variable $\tilde{t}=t/\epsilon = O(1)$. The Robin boundary condition at x=0 results from a linear potential profile in the SDL (due to the constant concentration) and resembles the classical boundary condition modeling external convection in heat or mass transfer,[40] where

$$\text{Bi}_e = \frac{L_e \cdot d_{sdl}}{p \cdot L_{sdl}} \tag{17}$$

plays the role of the Biot number. As shown in *Appendix B*, the exact solution can be expressed as a Fourier series,[41]

$$\phi(x,\tilde{t}) = \phi_m \cdot \sum_{n=0}^{\infty} \left( \frac{4\sin\lambda_n}{2\lambda_n + \sin 2\lambda_n} \right) e^{-\lambda_n^2 \tilde{t}} \cos(\lambda_n(x-1)) \tag{18}$$

where the discrete spectrum $\{\lambda_0, \lambda_1, \lambda_2, ...\}$ is defined by positive solutions of the transcendental equation, $\lambda_n \tan \lambda_n = \text{Bi}_e$. For $\text{Bi}_e$=2 (as in the numerical example in the next section) we obtain for $\lambda_i$: {1.08, 3.64, 6.58, …}. As shown below, a "diffusion layer" of charging (potential variation) spreads from the SDL interface across the electrode in a dimensionless time $\tilde{t}=t/\epsilon = O(1)$. After this initial penetration phase, the series solution (18) becomes much more useful, as it can be truncated after the first term to very high (exponential) accuracy.

As in the case of flat electrodes,[19] the linearized problem for low voltages or early times, Eq. (16), has a classical RC circuit interpretation,[25] that the porous electrode acts as a "transmission line" in series with the SDL "resistor", as shown in Fig. 2. This physical interpretation becomes clearer, if we briefly restore units. The dimensional time scale for the charging dynamics is

$$\tau_c = \frac{L_e^2}{D_e} \cdot \frac{\lambda_D^0}{h_p} = \left(\frac{\lambda_D^0 \cdot h_p}{D_e}\right) \cdot \left(\frac{L_e}{h_p}\right)^2 = \tau_{RC} \cdot N_{RC}^2 \tag{19}$$

where we recognize the "mixed" time scale $\tau_{RC} = \lambda_D^0 \cdot h_p / D_e$ (the geometric mean of the diffusion times across the pore and across the double layer) as the RC time constant[19] of a pore section (by analogy with parallel-plate electrodes separated by $h_p$) and $N_{RC} = \alpha$ as the effective number of such RC circuit elements along the length of a typical pore.

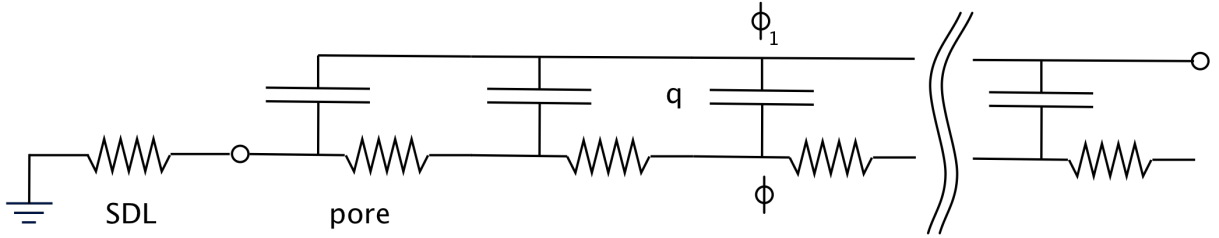

Fig. 2. Equivalent-circuit interpretation of the linearized charging dynamics of a porous electrode, Eq. (18), as an transmission line,[30] where the quasi-neutral solution in the pores acts as a series of resistors coupled to the electrode by parallel double-layer capacitors.

To improve on this analytical solution, a "weakly nonlinear" asymptotic approximation[19,22,24] can be systematically constructed in the asymptotic limit of thin double layers, as a regular perturbation series for $\tilde{t} = t/\epsilon = O(1)$. The leading-order approximation involves a uniform bulk concentration in Eq. (14)a, but without linearizing for small potentials. Instead, we obtain a nonlinear PDE for the leading-order potential in the supercapacitor regime:

$$\cosh\left(\frac{\phi_1 - \phi}{2}\right)\left(\frac{\partial \phi}{\partial \tilde{t}} - \frac{\partial \phi_1}{\partial \tilde{t}}\right) = \nabla^2 \phi \tag{20}$$

where we also allow for a time-dependent applied voltage $\phi_1(\tilde{t})$. The nonlinear prefactor in Eq. (20) is the (dimensionless) differential capacitance of the double layer in the GCS model,[19] and more generally, could be replaced by analogous expressions for many other quasi-equilibrium double-layer models (as reviewed in ref. 28).

In the GCS model, the differential capacitance, and thus the local "RC time", diverge exponentially with voltage, which causes the charging dynamics to slow down substantially as the local voltage leaves the linear regime. As a result the desalination regime (when the salt concentration starts to vary) is reached before the voltage much exceeds the thermal voltage, $|\Delta\phi| > 1$. This effect is reduced in modified double-layer models for finite-sized ions, which predict a decay of differential capacitance at large voltage, as the double layer expands due to ion crowding,[42,43] although this can be offset by other effects such as dielectric saturation.[28] Non-monotonic differential capacitance can lead to some surprising dynamical phenomena at blocking electrodes, such as flow reversal in AC electro-osmosis,[44] so it would be interesting to study the weakly nonlinear dynamics of porous electrodes in the supercapacitor regime using different double-layer models.

### III.C. Desalination Regime

For a sufficiently large applied voltage or small initial salt concentration (i.e. not very thin double layers inside the pores), the super-capacitor regime eventually transitions to the "strongly nonlinear" desalination regime, where the salt concentration within the porous electrode deviates significantly from its initial value. For flat electrodes, the nonlinear dynamics of this transition have been extensively analyzed.[19-24] Weakly nonlinear capacitive charging, Eq. (20), breaks down when the double layers adsorb a significant amount of neutral salt compared to that contained in the nearby bulk quasi-neutral solution. For smooth electrodes in the GCS model, the BTA criterion for validity of the weakly nonlinear dynamics in response to a suddenly applied DC voltage is[19]

$$\alpha_d = 4\sqrt{\frac{\epsilon}{\pi c}}\sinh^2\left(\frac{\Delta\phi_D}{4}\right) = 4\sqrt{\frac{\lambda_D(C)}{\pi h_p}}\sinh^2\left(\frac{ze\Delta\Phi}{4kT}\right) \ll 1 \tag{21}$$

Analogous frequency-dependent criteria can be derived for the breakdown of weakly nonlinear response to an AC voltage applied at flat electrodes.[24]

For the super-capacitor regime of a porous electrode, Equation (21) must hold at the scale of a single pore, and there may be additional constraints set by the spatially varying dynamics at the electrode length scale, as well as additional microscopic effects missing in our model, such as surface conduction (see below). The scalings in Eq. (21) show that the super-capacitor regime is valid for highly concentrated electrolytes (very thin double layers) and/or small voltages. Conversely, if the bulk ionic strength is large (as in CDI of seawater), the desalination regime requires a large applied voltage (>> $kT/ze$ with dimensions).

In the desalination regime, one must solve the full model (14) for $t = O(1)$, i.e. at the macroscopic diffusion time scale,

$$\tau_d = \frac{L_e^2}{D_e} = \frac{\tau_c}{\epsilon}. \tag{22}$$

At this time scale, the terms on the left hand sides of Eqs. (14)a and (14)b, representing the Maxwell current density and salt adsorption rate, are small, $O(\epsilon)$, but can become important if the voltage gets sufficiently large to violate Eq. (21).

For the GCS model, we can combine Eqs. (14)a and (14)b to obtain an alternate evolution equation for the concentration,

$$\left(1 - \frac{\epsilon}{\sqrt{c}}\tanh\frac{\phi_1 - \phi}{4}\right)\frac{\partial c}{\partial t} = \nabla^2 c + \epsilon\left(\tanh\frac{\phi_1 - \phi}{2}\right)\nabla\cdot(c\nabla\phi) \tag{23}$$

where the hyperbolic tangent factors tend to unity for highly charged regions, $|\Delta\phi| \gg 1$, in the desalination regime. For arbitrarily large voltages, in the asymptotic limit of thin double layers, the leading order behavior for $t = O(1)$ is again approximately governed by a linear diffusion equation

$$\frac{\partial c}{\partial t} = \nabla^2 c + O(\epsilon) \tag{24}$$

only now for neutral salt concentration, rather than the electrostatic potential. These reversed roles are clearly seen in the example of Fig. 3 below, where the potential at early times in (a) has a similar profile as the concentration at late times in (b). From Eq. (14) in this limiting regime, the potential is determined by the condition of small volume-averaged current into the double layers,

$$\nabla \cdot (c\nabla\phi) = O(\epsilon) \qquad (25)$$

which also determines the quasi-equilibrium volume-averaged profiles of charge, $q$, and neutral salt, $w$, adsorbed by the double layers.

The difficulty in using the simple limiting equations (24)-(25), however, is that the $O(1)$ initial conditions come from nontrivial asymptotic matching with the early-time super-capacitor regime. This requires solving the full equations numerically (as in the next section) or performing a more sophisticated asymptotic analysis, beyond the scope of this article. Similar problems relating to this dynamical transition are encountered for large DC voltages suddenly applied to smooth electrodes[19] or polarizable particles.[20] In contrast, the steady response to a large AC voltage removes this difficulty (while adding others) due to the imposed time scale of the AC period, which effectively selects one of the two limiting regimes.[24]

Finally, we note that a proper description of the desalination regime also requires including some additional physics that we have neglected. In addition to various electrochemical surface effects already mentioned, such as specific adsorption of ions and Faradaic reactions, there is another important source of nonlinearity in the ion transport equations, even for thin double layers. As noted by BTA,[19] when the condition (21) is violated, *surface conductivity* becomes comparable to the bulk conductivity, and thus the tangential transport in the double layer should also be considered.[20,31] This would be a major complication for volume averaging, which to our knowledge has never been considered in porous electrode theory.

**IV. Numerical Results**

As described in the previous section, the full model is difficult to solve, or even approximate, analytically when the applied voltage is large enough (or the initial double layers thick enough) to enter the desalination regime. In this section, we present numerical solutions of the full model for a typical case of CDI, which illustrate the two limiting regimes and the transition between them. In this section, calculation results are presented for the following parameter settings, namely $L_{sdl}=L_e=100$ μm, $d_{sdl}=D/D_e=1$. Furthermore, $\lambda_B=0.72$ nm, $C_\infty=10$ mM, $\lambda_D=3.0$ nm, porosity $p=0.5$, and $a=2\cdot10^7$ m$^2$/m$^3$, thus $\alpha=L_e/h_p=4000$, $\epsilon=\lambda_D^0/h_p=0.121$, and $\beta=8.41$. The effective thickness of the Stern layer, $\lambda_S$, is set to zero, thus $\Delta\phi_S=0$. We consider adsorption of cations in a negatively biased electrode ($\phi_1=-10$), and thus $i_e$, $j_{salt}$, $j_{charge}$ and $q$ are all positive numbers, while $\Delta\phi_d$ is of negative sign. All potentials are relative to the potential in bulk solution, just outside the SDL.

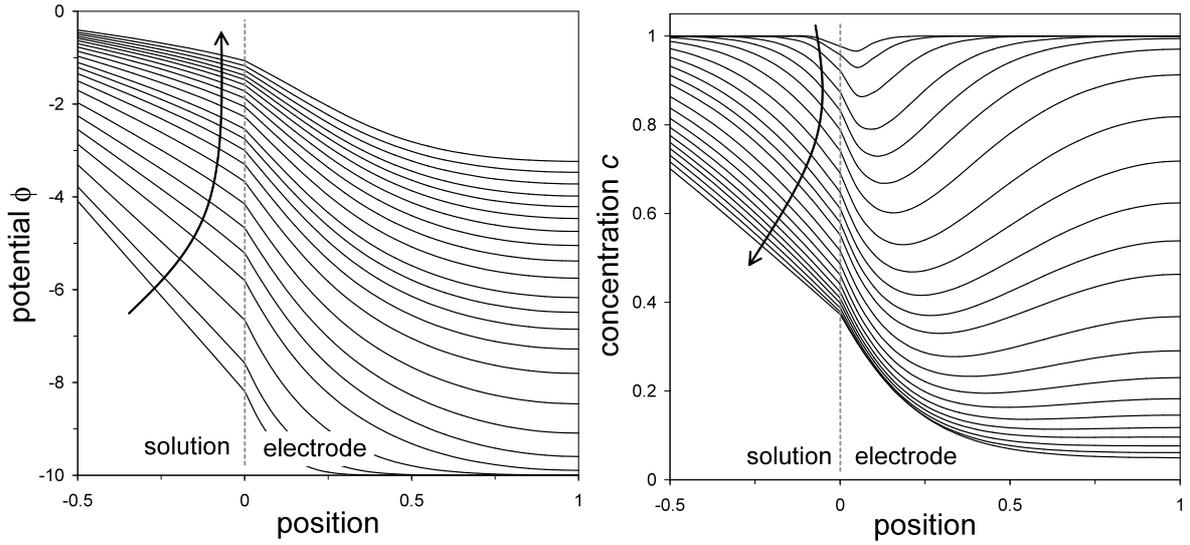

Fig. 3. Capacitive response of porous electrode to step change in electrode potential as function of time and position. Arrow shows the direction of time, 0.001<$t$<0.5. Parameter settings in the text. (a). Electrostatic potential, $\phi$. (b). Salt concentration, $c$.

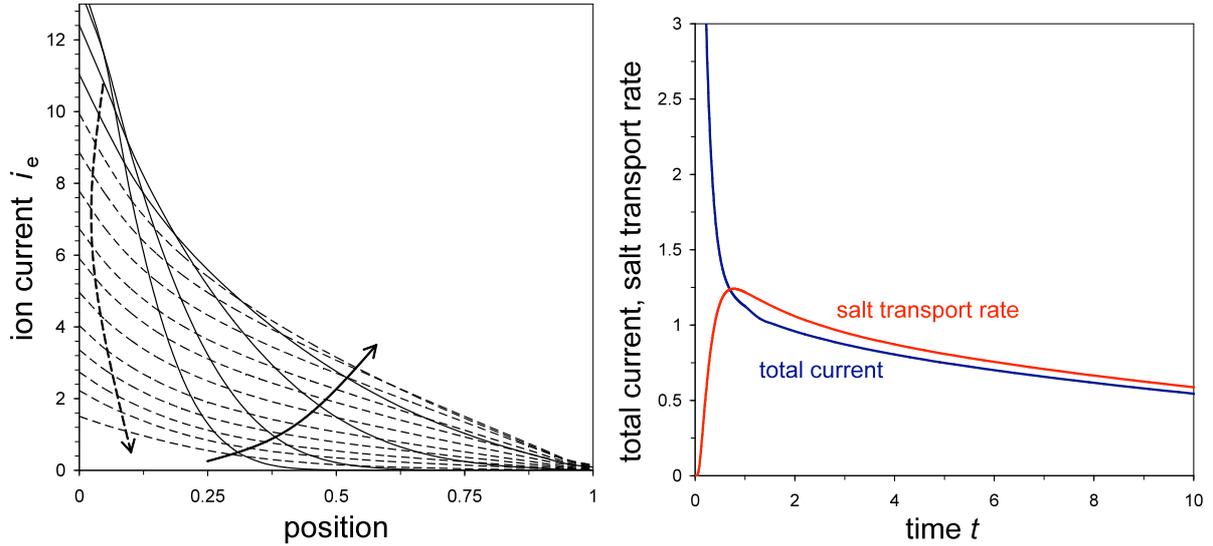

Fig. 4. (a). Ion current profiles in the electrode as function of time. Arrow shows the direction of time (0.001<$t$<0.5; first solid curves, next dashed curves). Ion current at position $x$=0 equals the total current. (b). Total current versus time $t$, and salt transport rate evaluated at the outer edge of the SDL.

Figs. 3 and 4 summarize our calculation results, where Fig. 3 shows the development of the electrostatic potential profile in time (Fig. 3a), as well as the salt concentration profile (Fig. 3b), while Fig. 4a shows how the ion current decreases across the electrode, and Fig. 4b how the measurable total current and salt adsorption rate (the latter evaluated at the edge of the SDL with bulk solution) change in time. In the calculation, upon applying the voltage, very rapidly charge is stored in the electrode, especially near its outer edge. Indeed, in Fig. 3a, even for the first curve at $t$~0.00125 ($\tilde{t}=0.0103$), the potential within the aqueous pore phase, $\phi$, is already significantly above the electrode matrix potential of $\phi_1$=-10, especially near the outer edge of the electrode. Note that for these initial times, the full numerical model is in perfect agreement with the supercapacitor-equation, Eq. (18). In this short time, quite some charge has already adsorbed at the internal electrode surface area, i.e., at the matrix/solution-interface, and the corresponding value of $\Delta\phi_D$ compensates the difference between $\phi_m$ and $\phi$. With increasing time progressively more charge adsorbs, $\Delta\phi_D$ increases, the profiles of potential $\phi$ in the aqueous pore phase gradually fade out, and eventually the pore potential $\phi$ becomes equal to the potential in the SDL and in bulk solution.

After application of the voltage, the ion concentration profiles in the electrode change more gradually. In Fig. 3b we clearly observe how first mainly near the outer electrode surface, salt is adsorbed from the SDL and from the pores in the electrode, to be adsorbed at the matrix/solution-interface. In the SDL slowly the classical steady-state profile is reached for which ion concentration linearly decays, while within the electrode the minimum in salt concentration shifts to the inner boundary. After $t\sim1$ gradually the concentration everywhere increases again [not shown] to finally reach unity again (i.e., equal to the concentration in bulk solution).

Fig. 4a shows how within the electrode the current carried by the ions gradually decreases with depth, to become zero at the inner electrode boundary. Indeed, the current carried by the electrons in the conductive matrix phase progressively increases in this direction such that the total current (which is equal to the ion current at $x$=0) remains constant. The total current is plotted as function of time in Fig. 4b together with the salt transport rate, which we evaluated at the outer edge of the SDL (divided by $p$). Though after an initial period both curves are close, they are certainly not overlapping, showing the relevance of distinguishing between charge transport (current) and salt transport.

The porous electrode-model can be incorporated in a larger-scale process model, e.g., for capacitive deionization, where the salt removal rate from bulk solution notably influences the bulk salt concentration, which in its turn influences the adsorption rates. Not only salt removal can be modeled, but it is just as well possible to model the subsequent step of salt release (after short-circuiting the cell). Though in this work we have applied the porous electrode-model to describe the capacitive charging of an electrode, Faradaic charge transfer processes including diffuse layer (Frumkin) effects[35-39,45-48] can also be included. For such Faradaic reactions, such as occur in the porous electrodes of fuel cells, it is the (local, and time-dependent) Stern layer potential difference (which is a function of the local charge density stored in the double layer at the matrix/solution-interface) which -together with the local ion concentration- self-consistently determines the charge-transfer rate (and is determined by it). Such a porous electrode model including Faradaic charge transfer relates to the work by Franco *et al.*[47] which is set up for a hydrogen fuel cell, where the proton is considered to be the only mobile species.

## V. Conclusion

We have presented a simple mathematical model for the nonlinear dynamics of charging and desalination by a porous electrode, which is biased by an applied voltage relative to the bulk solution. The model accounts for the transport of ions within the electrode pores and also in the solution phase outside the electrode. Both charge and neutral salt are stored in the double layers that form on the internal matrix/solution-interface within the electrode, which are assumed to the thin compared to typical pore dimensions. Using the classical Gouy-Chapman-Stern model for the double layer, analytical expressions are available for the local salt adsorption and charge density, which are essential elements of the porous-electrode model. The equations are combined into two PDEs for the volume-averaged salt concentration and electrostatic potential within the pores, and two dynamical regimes are identified: the "supercapacitor regime" at early times and/or small voltages and the "desalination regime" for late times and large voltages. Numerical calculations are presented for the salt concentration profiles as function of time, both within and outside the electrode region, as well as the profiles for potential, salt adsorption, charge density, and ion current. It would be straightforward to extend the theory for different models of the double layer, non-uniform porosity, or Faradaic reactions. A nontrivial extension would be to include geometry-dependent effect of tangential surface conduction within the pores and rigorously derive the strongly nonlinear, volume-averaged porous-electrode equations. It would also be interesting (and challenging) to relax the thin double-layer approximation for nano-porous electrodes. Applications include energy storage in supercapacitors and capacitive deionization by porous electrodes.

**Acknowledgments**

This work was supported by Voltea B.V. (Leiden, the Netherlands) (PMB) and by the National Science Foundation (USA) under contracts DMS-6920068 and DMS-0842504 and the MIT Energy Initiative (MZB). The authors also thank H. Bruning (Wageningen, the Netherlands) for useful suggestions.

**Appendix A: Formal derivation of porous electrode theory from the BTA equations**

For a binary electrolyte between blocking parallel-plate electrodes, BTA[19] used matched asymptotic expansions in the limit of thin double layers to derive effective boundary conditions

$$-\epsilon \frac{\partial q(\Delta\bar{\phi}_D, \bar{c})}{\partial \bar{t}} = \hat{n} \cdot (\bar{c}\bar{\nabla}\bar{\phi}) \tag{A1}$$

$$-\epsilon \frac{\partial w(\Delta\bar{\phi}_D, \bar{c})}{\partial \bar{t}} = \hat{n} \cdot \bar{\nabla}\bar{c} \tag{A2}$$

to be applied to the standard bulk equations for a quasi-neutral binary electrolyte,

$$0 = \bar{\nabla} \cdot (\bar{c}\bar{\nabla}\bar{\phi}) \tag{A3}$$

$$\frac{\partial \bar{c}}{\partial \bar{t}} = \bar{\nabla}^2 \bar{c} \tag{A4}$$

where bar accents indicate variables in the quasi-neutral bulk solution inside a pore (outside the thin double layers), length is scaled to $h_p$, time is scaled to $h_p^2/D$, and where $\Delta\bar{\phi}_D = \phi_1 - \bar{\phi}$ is the local diffuse-layer voltage. The effective boundary conditions (A1) and (A2) express conservation of charge and total salt, respectively, where $q$ is the integrated charge (per area) in the diffuse part of the double layer (and $-q$ is the surface charge density), and $w$ is the integrated excess total number of ions, relative to the nearby bulk solution. For multi-dimensional situations, the BTA equations (A1) and (A2) must be replaced by more general "surface conservation laws" with additional nonlinear terms describing tangential transport of ions in the double layers, derived in ref. 31 by matched asymptotic expansions in higher dimensions. We neglect such corrections, following all prior modeling on porous electrodes, although we believe they are important and should be considered in future work.

In order to *formally* perform volume averaging for a porous microstructure, we integrate (A3) and (A4) over the pore volume in a macroscopic volume element of (dimensionless) volume V (Fig. 5) and apply the divergence theorem to obtain surface integrals, with two contributions. The first contributions, from the double layers on the pores with (dimensionless) total area $A_p$, are evaluated using (A1) and (A2). The second contributions, from the quasi-neutral pore cross sections on the faces of the macroscopic volume element with (dimensionless) total area $A_v$, are related to the volume-averaged fluxes of charge and neutral salt. We relate the internal pore variables in (A1)-(A4) to the macroscopic, volume-averaged variables from the main text via the following definitions,

$$c = \frac{1}{V_p} \int_{V_p} \bar{c}\, d\bar{v} \tag{A5}$$

$$q(\Delta\phi_D) = \frac{1}{A_p} \oint_{A_p} q(\Delta\bar{\phi}_D)\, d\bar{a} \tag{A6}$$

$$w(\Delta\phi_D) = \frac{1}{A_p} \oint_{A_p} w(\Delta\bar{\phi}_D)\, d\bar{a} \tag{A7}$$

$$\nabla \cdot (c\nabla\phi) = \frac{D_e h_p^2}{D L_e^2} \frac{1}{V_p} \oint_{A_v} \hat{n} \cdot (\bar{c}\bar{\nabla}\bar{\phi})\, d\bar{a} \tag{A8}$$

$$\nabla^2 c = \frac{D_e h_p^2}{D L_e^2} \frac{1}{V_p} \oint_{A_v} \hat{n} \cdot \nabla \bar{c}\, d\bar{a} \qquad (A9)$$

which hold in the limit that the microscopic variables are slowly varying, and are roughly constant at the scale of individual pores. Recall that $h_p = V_p / A_p$ is the mean pore thickness, defined as the mean pore volume per pore area.

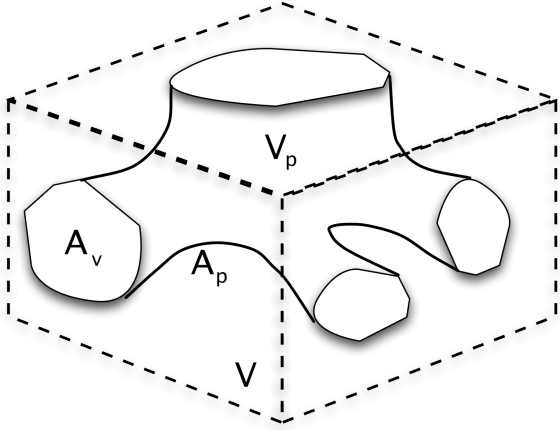

Fig. 5. Sketch of a macroscopic volume element of volume $V$ containing a pore space of volume $V_p$ and internal surface area $A_p$ which intersects the volume element with cross-sectional area $A_v$.

With these definitions, we arrive at two general, nonlinear PDEs for the volume-averaged response of porous electrodes,

$$\in \frac{\partial q(\Delta\phi_D, c)}{\partial t} = \nabla \cdot (c \nabla \phi) \qquad (A10)$$

$$\in \frac{\partial w(\Delta\phi_D, c)}{\partial t} = \nabla^2 c - \frac{\partial c}{\partial t} \qquad (A11)$$

which are analogous to the BTA boundary conditions (A1)-(A2) for locally smooth electrodes. These PDEs are consistent with classical formulations of porous electrode theory,[1-2] except that we have provided simple analytical expressions for the left hand sides and a formal derivation based on the asymptotic limit of thin double layers within the pores. In general, we could derive formulae for $q$ and $w$ for any concentrated-solution theory of the quasi-equilibrium double layer,[28] and this has been done for simple models of finite-sized ions[42] and applied to extend the BTA analysis of parallel-plate blocking electrodes.[43] The theory could also be naturally extended to multi-component, asymmetric electrolytes by replacing $q$ and $w$ with more general "surface concentrations" of the ionic species, given by integrals of the excess ionic concentrations (relative to the nearby bulk solution) across the diffuse part of the double layer.[31]

For the GCS model analyzed in the main text, the formula for $q$ is well known (and due to Gouy himself), and there is also a simple formula for $w$ (refs. 16,19)

$$q(\Delta\phi_D, c) = -2\sqrt{c} \sinh\frac{\Delta\phi_D}{2} \qquad (A12)$$

$$w(\Delta\phi_D, c) = 4\sqrt{c}\,\sinh^2\frac{\Delta\phi_D}{4} \tag{A13}$$

For the GCS model, the voltage drop is only over the diffuse part of the double layer, which is related to the full voltage drop, including the Stern layer, by a nonlinear capacitance equation.[19,38] The charge efficiency in the GCS model is given by,

$$\Lambda \equiv \frac{w}{q} = -\tanh\frac{\Delta\phi_D}{4} \tag{A14}$$

while the differential charge efficiency

$$\lambda \equiv \frac{j_{\text{salt}}}{j_{\text{charge}}} = \frac{\frac{\partial w}{\partial t}}{\frac{\partial q}{\partial t}} = \frac{\partial w}{\partial q} \tag{A15}$$

takes a simple form in the GCS-model,

$$\lambda\big|_c = -\tanh\frac{\Delta\phi_D}{2} \tag{A16}$$

with the additional constraint of a constant bulk salt concentration $c$. The latter expression (A16) appears in Eq. 106 of the "weakly nonlinear" analysis of BTA.[19] For "strongly nonlinear" analysis, when the salt concentration $c$ varies in time, the differential charge efficiency $\lambda$ becomes more complicated, and can be put the following form for the GCS model,

$$\frac{\partial q}{\partial t} = \frac{1}{\lambda}\frac{\partial w}{\partial t} + \frac{\Lambda}{\sqrt{c}}\frac{\partial c}{\partial t} = -\coth\frac{\Delta\phi_D}{2}\cdot\frac{\partial w}{\partial t} - \frac{1}{\sqrt{c}}\tanh\frac{\Delta\phi_D}{4}\cdot\frac{\partial c}{\partial t} \tag{A17}$$

which we used in deriving Eq. (23). We can also express Eq. (A17) as

$$\lambda = -\tanh\frac{\Delta\phi_D}{4}\left(1 - \frac{\Lambda}{\sqrt{c}}\frac{\frac{\partial c}{\partial t}}{\frac{\partial q}{\partial t}}\right) \tag{A18}$$

where the second term is missing in Eq. (4) in ref. 17. The two terms in Eq. (A17) or (A18) can be of comparable size at low voltage, but Eqs. (A13) and (A17) imply that the simple formula, $\lambda = \frac{j_{\text{salt}}}{j_{\text{charge}}} \approx -\tanh\frac{\Delta\phi_D}{2}$, becomes a good approximation in the desalination regime of large voltages.

**Appendix B: Analytical solution for linear response**

Here we derive the solution (18) of the linear initial-boundary-value problem (16) for the electrostatic potential in the supercapacitor regime, using the "Finite Fourier Transform" (generalized Fourier series) method.[40] The solution is expressed as a sum of separable terms,

$$\phi(x,t) = \sum_{n=0}^{\infty} \phi_n(t)\Phi_n(x)$$

where $\{\Phi_n\}$ are eigen-functions of the (self-adjoint) Laplacian operator satisfying the same homogeneous Robin boundary conditions

$$\frac{d^2\Phi_n}{dx^2} = -\lambda_n^2 \Phi_n, \quad \frac{d\Phi_n}{dx}(0) = Bi_e \Phi_n(0), \quad \frac{d\Phi_n}{dx}(1) = 0$$

which are orthogonal under the inner product, $\langle f,g \rangle = \int_0^1 f(x)g(x)dx$. If we scale the eigen-functions to be orthonormal, $\langle \Phi_n, \Phi_m \rangle = \delta_{n,m}$, then they take the form,

$$\Phi_n(x) = A_n \cos(\lambda_n(x-1)), \quad A_n^2 = \frac{4\lambda_n}{2\lambda_n + \sin 2\lambda_n},$$

where the eigenvalues are positive roots of $\lambda_n \tan \lambda_n = Bi_e$. Orthonormality implies $\phi_n = \langle \phi, \Phi_n \rangle$, which allows us to transform the PDE (12) into a simple ordinary differential equation,

$$\frac{d\phi_n}{dt} = \left\langle \frac{\partial \phi}{\partial t}, \Phi_n \right\rangle = \left\langle \frac{\partial^2 \phi}{\partial x^2}, \Phi_n \right\rangle = \left\langle \phi, \frac{d^2 \Phi_n}{dx^2} \right\rangle = -\lambda_n^2 \phi_n$$

using the self-adjoint property in the third step. The solution is

$$\phi_n(t) = A_n \phi_m \frac{\sin \lambda_n}{\lambda_n} e^{-\lambda_n^2 t}$$

since $\phi_n(0) = \langle \phi(x,0), \Phi_n(x) \rangle = \phi_m \langle 1, \Phi_n \rangle$.